# Improved Unet brain tumor image segmentation based on GSConv module and ECA attention mechanism

**Qiyuan Tian[1*], Zhuoyue Wang[2], Xiaoling Cui[3]**

[1]George Washington University, Washington, DC, USA.
[2]Department of Electrical Engineering and Computer Sciences, University of California, Berkeley, Berkeley, CA, USA.
[3]Independent Researcher, San Jose, CA, USA.
*Corresponding author email: chris.tqy128@outlook.com

**Abstract.** In this paper, an improved model of medical image segmentation for brain tumor is discussed, which is a deep learning algorithm based on U-Net architecture. Based on the traditional U-Net, we introduce GSConv module and ECA attention mechanism to improve the performance of the model in medical image segmentation tasks. With these improvements, the new U-Net model is able to extract and utilize multi-scale features more efficiently while flexibly focusing on important channels, resulting in significantly improved segmentation results. During the experiment, the improved U-Net model is trained and evaluated systematically. By looking at the loss curves of the training set and the test set, we find that the loss values of both rapidly decline to the lowest point after the eighth epoch, and then gradually converge and stabilize. This shows that our model has good learning ability and generalization ability. In addition, by monitoring the change in the mean intersection ratio (mIoU), we can see that after the 35th epoch, the mIoU gradually approaches 0.8 and remains stable, which further validates the model. Compared with the traditional U-Net, the improved version based on GSConv module and ECA attention mechanism shows obvious advantages in segmentation effect. Especially in the processing of brain tumor image edges, the improved model can provide more accurate segmentation results. This achievement not only improves the accuracy of medical image analysis, but also provides more reliable technical support for clinical diagnosis. To sum up, the improved U-Net model based on GSConv module and ECA attention mechanism proposed in this paper provides a new solution for brain tumor medical image segmentation, and its superior performance helps to improve disease detection and treatment effects, which is of great significance in related fields. In the future, we hope to further explore the application potential of this method in other types of medical image processing to advance the development of medical imaging.

## 1. Introduction

Brain tumor is one of the most common malignant tumors of the nervous system, and its early diagnosis and accurate segmentation are essential for developing effective treatment and evaluating prognosis. Traditional medical imaging methods, such as computed tomography (CT) and magnetic resonance imaging (MRI), play an important role in the detection and diagnosis of brain tumors [1]. However, manually analyzing these images is time-consuming and more likely to obtain subjective biases, which can lead to inconsistent results. Some studies have investigated the use of automatic image classification methods in the medical field [2], including applications specific to brain tumor analysis [3]. These approaches provide only a coarse-level diagnosis by focusing on the image as a whole, without capturing the finer details of its content. Consequently, automatic image segmentation technology has emerged as a significant research direction, offering a more detailed analysis by focusing on individual regions within the image.

Image segmentation aims to divide an image into multiple regions in order to better identify and analyze structures of interest. In the case of brain tumors, the goal is to accurately extract the tumor region and its boundaries so that doctors can make more precise assessments and interventions [4]. Deep learning algorithms are particularly suited to processing complex data patterns, which makes them excellent in the field of medical image analysis [5]. The convolutional neural network (CNN) is one of the most widely used architectures in deep learning, capable of effectively extracting features through multi-layer convolutional operations. It has demonstrated outstanding performance in classification and regression tasks across various critical national industries, including infrastructure [6], defense [7], and financial markets [8]. For brain tumor image segmentation, CNN can automatically learn to extract features from the original image to achieve efficient and accurate segmentation.

In recent years, some network architectures specially designed for medical image segmentation tasks, such as U-Net[9] and SegNet[10], have been widely used in brain tumor image segmentation. These networks effectively retain spatial information by introducing mechanisms such as jump connections and upsampling, thereby improving the ability to capture detailed parts such as tumor boundaries. For example, U-Net structure consists of encoder and decoder, which realizes feature extraction and reconstruction through layer-by-layer down-sampling and up-sampling, which greatly improves the performance of medical image segmentation task [11].

In short, brain tumor image segmentation, as a challenging medical image processing task, is moving towards more automation, high efficiency and precision with the development of deep learning technology. This not only helps to improve the efficiency of clinical diagnosis, but also provides better treatment and prognosis assessment for patients. In this paper, the most commonly used Unet algorithm in brain tumor medical image segmentation was selected, GSConv module and ECA attention mechanism were introduced to improve and optimize the model, and the segmentation effect was improved.

**2. Data set source**
Brain images from Kaggle open source data set, data set web site (https://www.kaggle.com/datasets/nikhilroxtomar/brain-tumor-segmentation), The data set contains two folders: images and masks. 500 images of image and mask were selected for the experiment. Part of the data set is displayed, as shown in Figure 1. The top is the original image, and the bottom corresponds to the segmentation results of the brain tumor part.

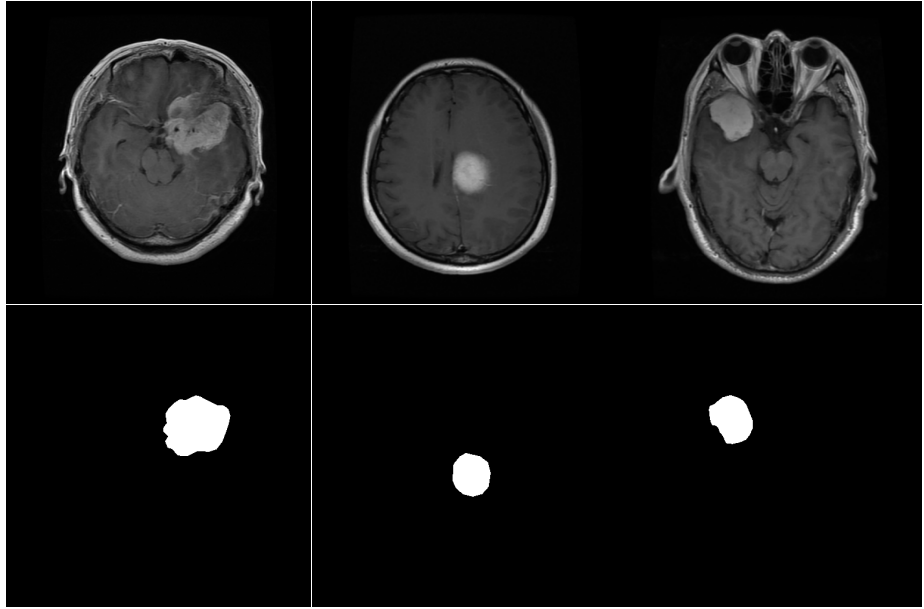

**Figure 1.** Part of the data set.

### 3. Image preprocessing
Firstly, the image is converted to gray level image, and histogram equalization is carried out. Through equalization processing, the details in low-contrast images can be significantly improved, making the features more obvious. In this paper, the image is preprocessed according to the following steps:

    1. Calculate the histogram: calculate the frequency of each gray level in the gray image to form the histogram. This step helps you understand the distribution of individual brightness values in the current picture.

    2. Calculate the Cumulative distribution function (CDF) : Calculate the cumulative distribution function based on the histogram, which helps us to know the total frequency of all levels before each gray level, so as to prepare for subsequent mapping.

    3. Normalized CDF: Map the CDF to the range [0, 255] so that it can cover the entire available grayscale space.

    4. Remapping gray value: According to the normalized CDF, the original gray value is mapped to the new value to generate a new equalized gray image.

### 4. Method

*4.1. U-net*

    U-net is a deep learning model for image segmentation, its structure is characterized by a symmetrical "U" shape design, including encoder and decoder parts. The schematic diagram of the Unet model is shown in Figure 2.

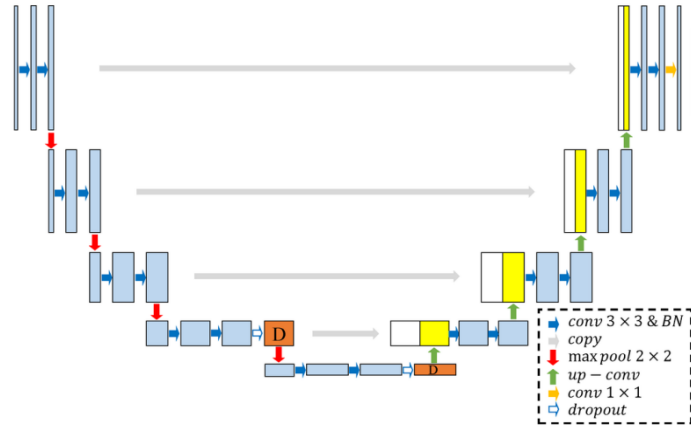

**Figure 2.** The schematic diagram of the Unet model.

The left half of the U-Net is the encoder, consisting of a series of convolutional layers and pooling layers. By gradually reducing the spatial dimension of the feature map and increasing the number of feature channels, advanced features in the image are extracted. This process enables the model to capture a wider range of context information [12].

The right half is the decoder that restores the feature map to the dimensions of the original input image by deconvolution (or upsampling). After each up-sampling step, the decoder combines information from the corresponding layer of the encoder, and this jump connection helps to preserve the details, making the segmentation result more refined.

U-Net usually uses cross entropy loss or Dice coefficient as optimization objective to improve segmentation accuracy. In this way, the model can effectively learn the boundary between the foreground and the background.

*4.2. ECA attention mechanism*

ECA attention mechanism is a channel attention mechanism for improving the performance of convolutional neural networks (CNNS), designed to enhance the model's focus on important features by effectively recalibrating feature channels. The model schematic diagram of ECA's attention mechanism is shown in Figure 3. The core idea of ECA's attention mechanism is to generate channel weights through local context information rather than global average pooling [13]. This process can be broken down into the following steps:

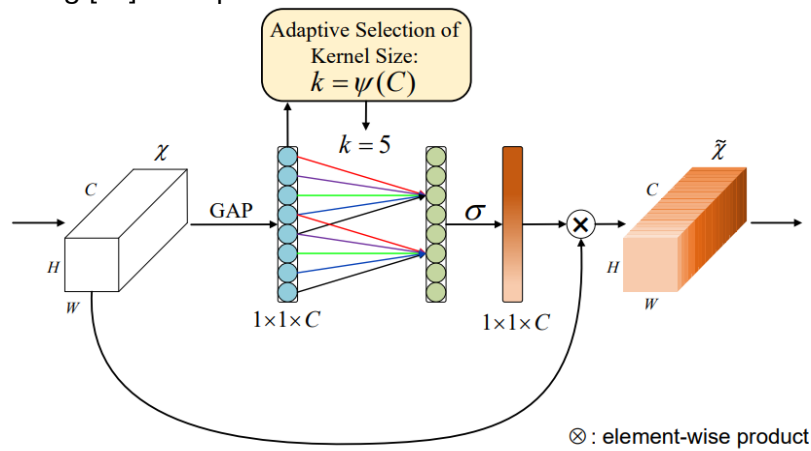

**Figure 3.** The model schematic diagram of ECA's attention mechanism.

1. Feature extraction: First, extract the features of each channel from the input feature map.

2. Local convolution operation: Unlike global average pooling, ECA uses one-dimensional convolution to capture the relationship between adjacent channels. This method can make full use of local context information while maintaining computational efficiency.

3. Generate weights: After the convolution operation, a weight vector will be obtained. This vector is normalized by the Sigmoid activation function to ensure that the ownership revalue is between 0 and 1.

4. Weighted feature maps: Finally, the generated weights are applied to the original feature maps, reinforcing important channels while suppressing unimportant ones by multiplying elements by elements.

In conclusion, ECA attention mechanism, with its high efficiency and superiority, provides a new way of thinking for deep learning models and helps to improve the ability of feature representation.

*4.3. GSConv*

GSConv is a novel convolution operation designed to improve the performance and efficiency of deep learning models. Its core principle is a combination of grouping convolution and displacement mechanisms [14]. First, GSConv divides the input feature graph into multiple groups and performs convolution operations for each group independently, which can significantly reduce the number of parameters and computational complexity while maintaining the expressiveness of the model. Then, during the convolution of each grouping, a displacement operation is introduced, that is, some channels in the feature map are moved left or right to capture more local context information. This mechanism enhances the model's understanding of spatial structure, thus improving the diversity of feature extraction.

Finally, after displacement, the feature map will be reassembled to integrate information from different groups by adding or splicing elements, so that the model can integrate information from multiple directions and further improve the overall performance. GSConv is not only computationally more efficient than traditional fully connected convolution, which is suitable for resource-constrained devices, but also makes the model perform better when dealing with complex patterns by enhancing feature learning ability. Therefore, GSConv provides a flexible and powerful feature extraction method for deep learning, which helps to promote the development of various visual tasks [15].

*4.4. Improved on GSConv module and ECA attention mechanism*

GSConv enhances feature extraction with packet convolution and displacement operations. In each convolutional layer, the input feature maps are divided into multiple groups and processed independently, which not only reduces the computational complexity, but also improves the sensitivity of the model to local features. The displacement operation further improves the ability of the model to capture spatial information, so that the network can better understand the complex structure, and thus improve the segmentation effect.

Combining these two technologies, the improved U-Net is able to extract and utilize multi-scale features more efficiently while flexibly focusing on important channels, significantly improving performance in medical image segmentation tasks. This combination makes the model more robust in processing complex images and helps to achieve higher quality segmentation results.

**5. Result**

In the experimental Settings, the epoch is set to 50, the input picture is [512,512], the learning rate is set to 0.0001, the CPU is 32G, the python version is 3.8, based on the pytorch framework, and the graphics card is 3090. In terms of data set partitioning, the data set is divided into training set and verification set according to the ratio of 4:1, that is, 80% of the data is used for training, 20% of the data is used for verification, and 8 images are randomly selected for testing.

In terms of experimental parameters, this paper uses loss to record the training process and output the loss change curve of the training set and the verification set. Loss function is a concept in machine learning and deep learning, which is used to quantify the gap between the predicted value and the real value of the model. By minimizing the loss function, the model can continuously optimize its parameters during training, thereby improving prediction accuracy. In this paper, miou is used to evaluate the segmentation effect of the model. miou is an effective and simple method to evaluate the performance of semantic segmentation models. By evaluating individual classes, it can provide information about how the model performs in different situations.

First, the Unet model is used for training, and loss change curves of training set and test set are output, as shown in Figure 4.

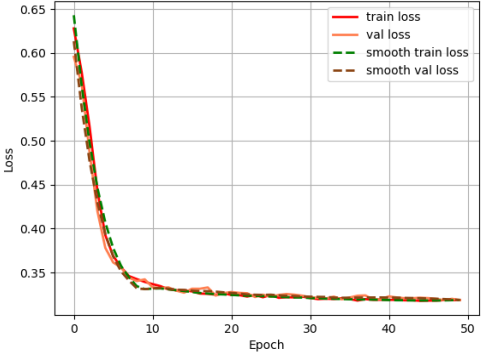

**Figure 4.** The loss change curves.

According to the changes of the loss curves of the training set and the test set of the Unet model, the loss value of the training set and the test set drops to the lowest point after the 8th epoch, and the loss value gradually converges and becomes stable.

The miou change curve of the training set is output, and the result is shown in Figure 5.

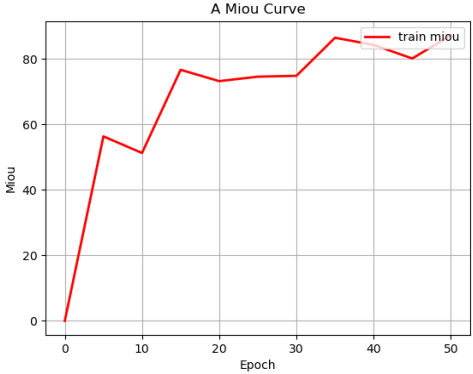

**Figure 5.** The miou change curve of the training set.

According to the miou curve, after the 35th epoch, miou gradually approaches 0.8 and remains stable.After experiments with Unet, we used an improved Unet model based on GSConv module and ECA attention mechanism to segment brain tumor images. loss change curves of training set and test set are output, as shown in Figure 6.

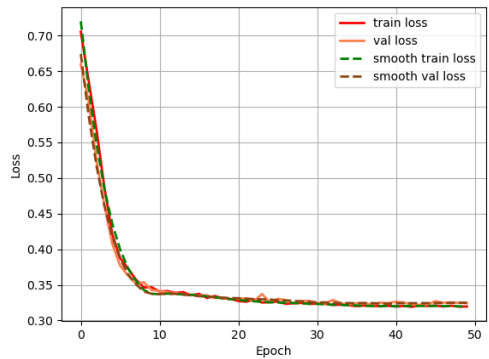

**Figure 6.** The loss change curves.

Based on the changes of the loss curves of the training set and test set of the improved Unet model based on the GSConv module and ECA attention mechanism, it can be seen that the loss value of the training set and test set drops to the lowest point after the 8th epoch, and the loss value gradually converges and becomes stable.

The miou change curve of the test set is output, and the result is shown in Figure 7.

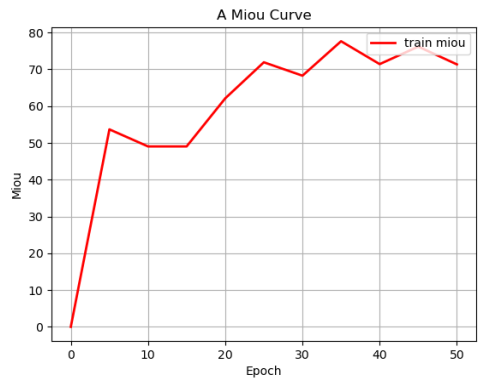

**Figure 7.** The miou change curve of the test set.

According to the miou curve, miou gradually becomes stable after the 30th epoch.

After the end of the two groups of experiments, the trained weights were respectively used for testing, the results of model testing were output, and the effect of the two groups of models on brain tumor image segmentation was compared, as shown in Figure 8.

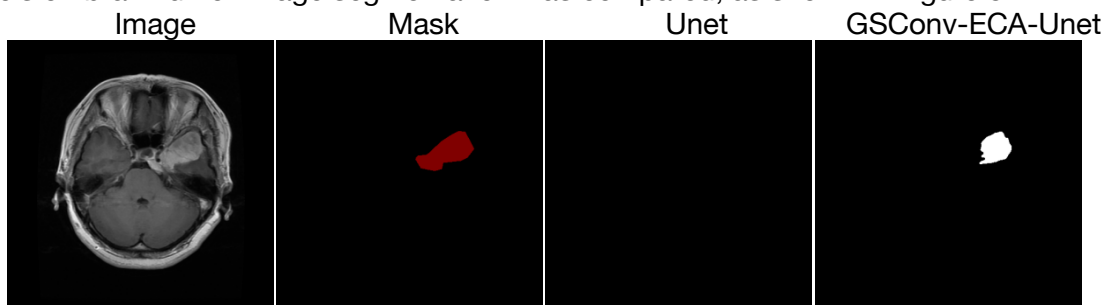

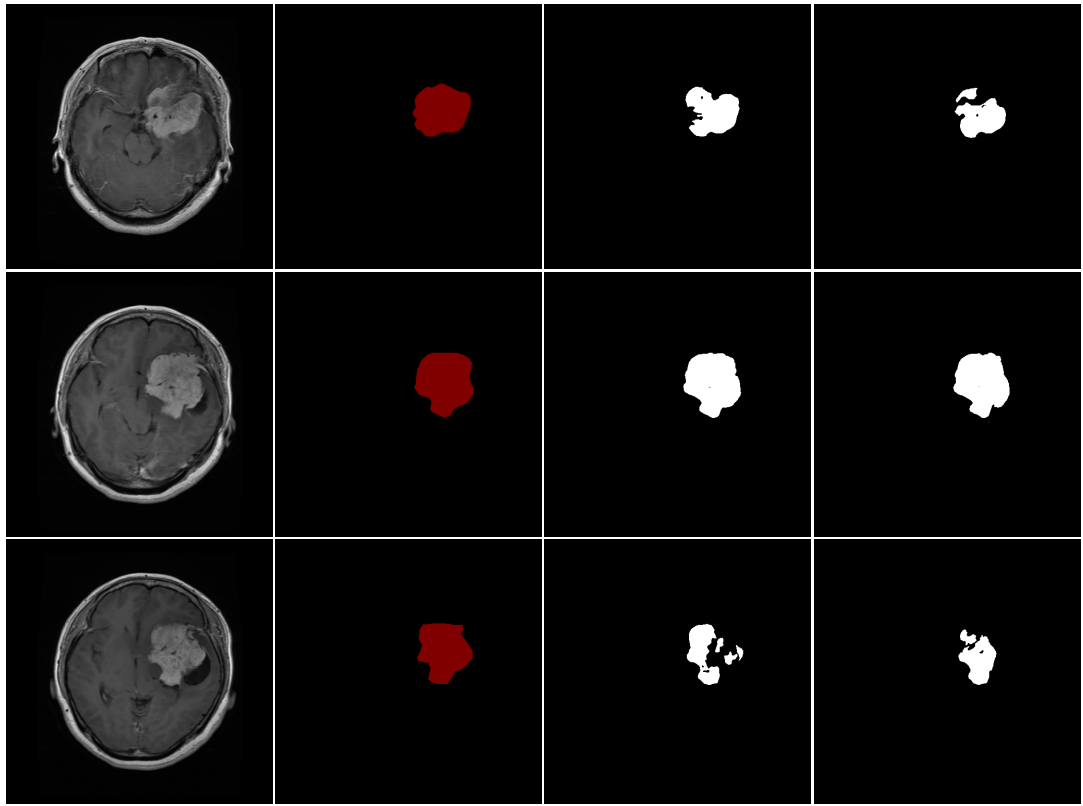

**Figure 8.** The effect of the two groups of models on brain tumor image segmentation.

According to the segmentation results of the test set, it can be seen that the improved Unet model based on GSConv module and ECA attention mechanism has better segmentation effect than Unet, especially in the edge segmentation effect. The improved Unet model based on GSConv module and ECA attention mechanism proposed in this paper shows excellent segmentation effect.

6. **Conclusion**

In this paper, we adopt U-Net algorithm for brain tumor medical image segmentation, and introduce GSConv module and ECA attention mechanism on this basis to improve and optimize the model. Through these innovations, we have successfully improved segmentation, enabling the improved U-Net to extract and utilize multi-scale features more efficiently while flexibly focusing on important channels. This method significantly enhances the performance of the model in medical image segmentation tasks, especially in the processing of complex brain tumor structures.

From the results observed in the training process, the improved U-Net model shows good convergence in the change of the loss curve of the training set and the test set. After the eighth epoch, the loss value quickly fell to its lowest point and gradually stabilized. This indicates that the model can effectively capture important features in the data during the learning process, thereby reducing the risk of overfitting. At the same time, by calculating the change curve of the mIoU, we find that after the 35th epoch, the mIoU value gradually approaches 0.8 and remains stable. This further proves the effectiveness of our proposed improvement scheme in improving model performance.

It is worth mentioning that through comparative experiments, we found that the improved U-Net model based on GSConv module and ECA attention mechanism is significantly better than the traditional U-Net in terms of segmentation effect, especially in the edge region segmentation ability. The segmentation results on the test set clearly demonstrate the accuracy and robustness of the model in processing brain tumor images. This not only helps to improve the accuracy of clinical diagnosis, but also provides more reliable data support for subsequent treatment plans.

In summary, the U-Net model proposed in this study, based on GSConv module and ECA attention mechanism improvement, provides a novel and efficient method for brain tumor medical image segmentation. This achievement not only promotes the development of medical image processing technology, but also helps to improve the efficiency of patient diagnosis and treatment. In the future, we hope to extend this method to other types of medical image analysis in order to achieve a wider range of applications. Through continuous optimization and improvement of the model, it is believed that it can bring more positive effects to the medical field and contribute to the development of precision medicine.